\documentclass[11pt]{article}
\usepackage{amssymb,amsmath,amsbsy}
\def\beq{\begin{eqnarray}}
\def\eeq{\end{eqnarray}}
\def\bea{\begin{eqnarray*}}
\def\eea{\end{eqnarray*}}

\def\centeron#1#2{{\setbox0=\hbox{#1}\setbox1=\hbox{#2}\ifdim
\wd1>\wd0\kern.5\wd1\kern-.5\wd0\fi
\copy0\kern-.5\wd0\kern-.5\wd1\copy1\ifdim\wd0>\wd1
\kern.5\wd0\kern-.5\wd1\fi}}
\def\ltap{\;\centeron{\raise.35ex\hbox{$<$}}{\lower.65ex\hbox{$\sim$}}\;}
\def\gtap{\;\centeron{\raise.35ex\hbox{$>$}}{\lower.65ex\hbox{$\sim$}}\;}

\def\singleandthirdspaced{\baselineskip=\normalbaselineskip\multiply
    \baselineskip by 130\divide\baselineskip by 100}

\def\dslash{\not{\hbox{\kern-2pt $\partial$}}}
\def\Dslash{\not{\hbox{\kern-4pt $D$}}}
\def\Oslash{\not{\hbox{\kern-4pt $O$}}}
\def\Qslash{\not{\hbox{\kern-4pt $Q$}}}
\def\pslash{\not{\hbox{\kern-2.3pt $p$}}}
\def\kslash{\not{\hbox{\kern-2.3pt $k$}}}
\def\qslash{\not{\hbox{\kern-2.3pt $q$}}}

\newcommand{\newc}{\newcommand}
\newc{\qbar}{{\overline q}}
\newc{\Kahler}{K\"ahler }
\newc{\deltaGS}{\delta_{\rm GS}}
\begin{document}
\begin{titlepage}
\begin{flushright}
{\large hep-th/yymmnnn \\ SCIPP-2009/14\\NHETC-2009-19
}
\end{flushright}

\vskip 1.2cm

\begin{center}

{\LARGE\bf Deriving particle physics from quantum gravity:\\ a plan}

\vskip 1.4cm

{\large Tom Banks}
\\
\vskip 0.4cm

 {\it Department of Physics and SCIPP\\
 University of California, Santa Cruz, CA 95064\\
 E-mail: {banks@scipp.ucsc.edu}\\
 {\it and}\\
 Department of Physics and NHETC, Rutgers University\\
 Piscataway, NJ 08540 }

 \vskip 4pt

\vskip 1.5cm

\begin{abstract}
I give a short review of the holographic approach to quantum
gravity, with emphasis on its application to deriving the properties
of elementary particles.

\end{abstract}

\end{center}

\vskip 1.0 cm

\end{titlepage}
\setcounter{footnote}{0} \setcounter{page}{2}
\setcounter{section}{0} \setcounter{subsection}{0}
\setcounter{subsubsection}{0}

%%%%%%%%%%%%%%%%%%%%%%%%%%%%%%%%%%%%%%%%%%%
%%%%%%%%%%%%%%%%%%%%%%%%%%%%
\singleandthirdspaced

\section{Introduction}

The holographic approach to the quantum theory of gravity is an
attempt to generalize existing string models of gravity in order to
describe cosmology and real world particle physics, including
supersymmetry breaking.  Its basic starting point is a strong form
of the Fischler-Susskind-Bousso covariant entropy conjecture, which
identifies the area in Planck units\footnote{We will use the
notation $G^{-\frac{1}{2}} =  M_P \approx 10^{19} $ GeV $ =
\sqrt{8\pi} m_P$, and units where $\hbar = c = 1$.} of the
holographic screen of a causal diamond, with four times the
logarithm of the dimension of the Hilbert space describing all
possible observations in that diamond. When combined with the usual
connection between causally separated regions and commuting operator
algebras, it provides a way to encode all of the data in a
Lorentzian geometry in terms of purely quantum mechanical concepts.

A quantum space-time is a collection of Hilbert spaces with a set of
causal relations between them.  Each Hilbert space is associated
with a causal diamond.  A common tensor factor of two Hilbert spaces
(a unitary map from a tensor factor of one to a tensor factor of the
other) is identified with the maximal area causal diamond in the
space-time overlap between the two diamonds. A pattern of overlaps
determines the geometry. The Hamiltonian dynamics within each
Hilbert space is constrained by conditions of compatibility with
that on overlapping spaces.  It may be that all such compatible nets
of quantum systems, with the property that Hilbert spaces become
very large in the future, generate Lorentzian geometries satisfying
Einsteins' equations coupled to some sort of matter.

In this holographic description of quantum gravity, the space-time
metric is not a fluctuating quantum variable. That is to say, the
causal structure and conformal structure of the emergent space-time,
come from exact features of the quantum theory: tensor inclusions
and Hilbert space dimensions. In this view the space-time metric
should be thought of as a collective variable, somewhat analogous to
invariants in large $N$ expansions. In perturbation theory, the
variable satisfies some kind of classical equations, and instanton
solutions of these equations can even be used to compute tunneling
probabilities.  However there is no sense in which the theory can be
viewed non-perturbatively as a quantization of these equations or a
path integral over fluctuating configurations of the metric.

\subsection{The {\bf D}ense {\bf B}lack {\bf H}ole {\bf F}luid}

There is only one solved example which supports the
conjecture that this sort of quantum system gives rise to an emergent space-time\cite{holocosm}, and it corresponds to a flat FRW
cosmology, with equation of state $p = \rho$, which saturates the
covariant entropy bound at all times.  It is called the dense black
hole fluid (DBHF).  Heuristic arguments\cite{holocosm2} show that
local low entropy defects in the DBHF can evolve as more ordinary
cosmologies, but must evolve to an asymptotic de Sitter (dS)
universe in the future, with cosmological constant (c.c.) determined
by a cosmological initial condition.  It is related by the
holographic principle to the dimension of the Hilbert space of the
subsystem that began in a low entropy state.

I will include a brief presentation of the DBHF here. To describe
the Hilbert space of a general observer in a Big Bang cosmology, we
start from the observation that in the classical geometry, the
observer's particle horizon shrinks to zero near the Big Bang, and
expands with the universe (we will not be describing Big Crunch
cosmologies here). The holographic connection between area and
entropy, and the translation of causality into commutation of
operators says that we should describe this quantum mechanically as
a sequence of Hilbert spaces, ${\cal H}_n$, with ${\cal H}_n = {\cal
H}_{n-1} \otimes {\cal P}$. Since our present discussion will lead
to an open subluminally expanding universe, $n$ will eventually go
to $\infty$ and we call the limiting Hilbert space ${\cal
H}_{\infty}$.  The single pixel Hilbert space ${\cal P}$ describes
all of the degrees of freedom that can be observed on a holographic
screen {\it in the non-compact dimensions of space-time}\footnote{It
is worth noting that in the holographic context, a future
asymptotically dS universe has a non-compact spatial topology. The
maximal causal diamond of an observer in such a space-time has a
null boundary, whose holographic screen has finite area.}. In a
later section I will argue that it is a representation of a
super-algebra, which encodes the geometry of the compact dimensions.
As a consequence of the holographic principle, that geometry must
have only a finite dimensional function algebra. A natural framework
to describe it is finite dimensional non-commutative geometry, also
called fuzzy geometry. ${\rm ln\ dim\ }{\cal P}$ is one quarter of the area of the smallest pixel of holographic screen, measured in units of the $d$ dimensional Planck scale.  $d$ is the number of non-compact dimensions.

The dynamics of our fiducial observer is described by a sequence of
unitary transformations $U(k)$, all operating in the full space
${\cal H}_{\infty}$ . If we postulate that $U(k) = U_{in} (k)
\otimes U_{out} (k) $ where $U_{in} (k)$ operates in ${\cal H}_k$,
then we incorporate the notion of {\it particle horizon} into our
holographic cosmology: the degrees of freedom in ${\cal H}_k$ do not
interact with the rest of the universe until time $k + 1$ , when
"the particle horizon has grown large enough to incorporate more
degrees of freedom".

All of experimental physics, throughout the history of the universe,
can be viewed as ``the observations of a single observer" , but we
like our theoretical structures to incorporate multiple observers,
related by consistency conditions.  In the present framework we do
this by introducing a lattice, which encodes the topology of the
non-compact dimensions of space. This topology does not change with
time. We will let ${\bf x}$ symbolize points on the spatial lattice
and ${\cal H}_n ({\bf x}) $ be the Hilbert space at time $n$ of the observer
whose world line is labeled by ${\bf x}$ . From the bulk
space-time point of view, we are in a synchronous coordinate system,
defined by {\it equal area time slicing}: the area of the holographic screen of the causal diamond stretching from the observer's current position back to the Big Bang, is the same at every point on the lattice. It should be noted that the integer $n$ does not directly count the time, but rather the area of the causal diamond.  In a wide variety of space-time geometries, $n$ is related to the proper time coordinate of ``natural" observers by $n \sim t^{d-2}$.

For the present discussion, we will ignore compact dimensions.
Anticipating the discussion of section 2, we conclude that the
pixel Hilbert space, ${\cal P}$, is the irreducible representation
of $$[S_a , S_b ]_+ = \delta_{ab},$$ where $S_a$ are the real
components of the transverse ($d - 2$ dimensional) spinor.

To complete the definition of a general quantum cosmology we must
specify the overlap Hilbert space ${\cal O}_n ({\bf x, y})$, which
is a tensor factor of both ${\cal H}_n ({\bf x})$ and ${\cal H}_n
({\bf y})$, for each pair of points on the lattice. By definition (of nearest neighbor), for nearest neighbor points we have ${\cal O}_n ({\bf x, x_{n.n.}}) = {\cal P}$.  Furthermore, the overlap must decrease in dimension as the minimum number of lattice steps between the two points
increases. We will see that it is this decrease which defines the
geometrical distance function on the lattice.

Geometrically, the overlap Hilbert space is associated with the
largest area causal diamond that fits in the intersection between
the causal diamonds represented by ${\cal H}_n ({\bf x})$ and ${\cal
H}_n ({\bf y})$. The geometrical interpretation is only supposed to
make sense when all Hilbert space dimensions are large. It's clear
that the causal structure and conformal factor of the emergent
space-time geometry are completely fixed by the quantum rules.
Geometry does not fluctuate in this model of quantum gravity. It is
an emergent classical variable, morally similar to the singlet
variables of large $N$ matrix models.

There is a very strong set of constraints on the quantum dynamics of
a holographic cosmology. Each sequence of overlap Hilbert spaces
${\cal O}_n ({\bf x, y})$ has two sequences of unitary
transformations induced on it from the dynamics on the individual
observer spaces. These two sets of transformations must be unitarily
related
$$U(n, {\bf x};{\bf  x,y}) = W^{\dagger} (n, {\bf x,y}) U(n, {\bf y}; {\bf  x,y}) W (n, {\bf x,y}) .$$

The definition of the DBHF dynamics is motivated by the observation
of Fischler and Susskind\cite{fs} that a flat FRW universe with
equation of state $p = \rho$ can saturate the covariant entropy
bound at all times. We thus search for a random Hamiltonian on the
individual observer's Hilbert space. Begin by choosing ($U(n) \equiv
e^{ i H(n)}$)

$$H(n) = \sum S_a (K) A(n; K,M) S_a (M) ,$$ where $A(n; K,M)$ is a random $n\times n$ antisymmetric matrix chosen (independently at each n) from the Gaussian ensemble. It is well known that as $n \rightarrow \infty$, $H(n)$, as a consequence of the Wigner semi-circle law, approaches a Hamiltonian for free massless relativistic $1 + 1$ dimensional fermions, with a cutoff.  If we now add a random polynomial of order $\geq 6$ in the $S_a (K)$ to $H(n)$, the large $n$ behavior is unchanged, away from the cutoff scale, because of the standard $1 + 1$ dimensional renormalization group flows. The thermodynamic quantities obey scaling laws in the limit, because the fixed point is scale invariant.  Random changes in $A(n, K,M)$ as $n$ is varied,
guarantee that the system will explore its full Hilbert space, given
a generic initial state. The covariant entropy bound is thus
saturated.

Define the overlap Hilbert spaces by ${\cal O}_n ({\bf x,y}) = {\cal
H}_{n - d({\bf x,y})},$ where $d({\bf x,y})$ is the minimum number of
lattice steps between the points.  When the subscript becomes
negative ${\cal O}$ vanishes. We can define a {\it causal boundary}
between two points at any ``time" $n$, by looking at the locus of
points beyond which the overlap vanishes.  For large $n$, this
occurs at large values of $d({\bf x,y})$, and on any regular lattice
with the topology of flat space the locus approaches a sphere. Thus,
for large $n$ the system has an emergent spherical symmetry. 

We can satisfy the infinite set of dynamical consistency conditions
by insisting that the random Hamiltonian seen by each observer at a
given time is independent of ${\bf x}$ and letting the dynamics in
the overlaps be exactly the dynamics seen by an observer at the earlier 
time, whose causal diamond has the same size as the overlap. This is probably the only way to satisfy the consistency conditions given the random dynamics for an individual observer.

We have thus proven that the geometry that emerges from this system
is homogeneous and isotropic.  The fact that it is flat follows from
the scale invariance of the large $n$ dynamics. The flat FRW metric,
with single component equation of state, has a conformal Killing
vector corresponding to simultaneous scaling of space and time
coordinates, with different exponents: 
$$ ds^2 = - dt^2 + t^a d({\bf x})^2 ,$$
$$t \rightarrow \lambda t \ \ \ {\bf x} \rightarrow \lambda^{1 - \frac{a}{2}} {\bf x} .$$
We identify this with the
emergent quantum scale invariance and learn that in this particular
case (unlike, for example a gas of photons)the conformal isometry is
in fact a symmetry of the full quantum system.   Negatively curved
FRW space, which is compatible with the topology of our lattice,
would not have such a scaling symmetry.

The fact that the equation of state of this system is $p = \rho$
already follows from the fact that its thermodynamics is that of a
conformal field theory in $1 + 1$ dimensions.  We define the space
time energy density of the FRW model to be the thermal expectation
value of the $1+1$ dimensional energy density. There are a variety
of other checks of the flatness, homogeneity, and isotropy of the
emergent geometry, and of its equation of state. These have been
explained in \cite{holocosm}.

The DBHF is not a nice place to visit, and you couldn't live there
if you tried. All the degrees of freedom within a horizon volume are
in equilibrium at all times.  Heuristically, the universe consists
at all times of a single black hole which fills the particle
horizon.  Fischler and I proposed to model a more realistic
cosmology by imagining a sprinkling of low entropy regions at the
time of the Big Bang, which form a defect in the DBHF. This defect
has a fixed size and shape in the geometry defined by the DBHF. The
size and shape are determined by maximizing the initial entropy
subject to the constraint that a more or less normal region of the
universe survives into the future. We do not yet know much about the
shape, except that it cannot be even approximately spherical.  The
size is a cosmological initial condition, which is fixed only by
insisting on properties of the asymptotic future state. We argue
that this initial condition determines the c.c.   The point is that
the quantum meaning of {\it the size of the initial defect} is just
the logarithm of the number of quantum states associated with the
initial defect region.  This number is finite.  Given the model of
quantum dS space in section 3, as a quantum system with a finite
number of states, it seems clear that these two numbers are the
same.

A simple calculation in GR confirms this identification.  Consider a
sphere of normal FRW universe, embedded in the DBHF and ask whether
the causal diamond of a normal observer can last indefinitely. The
Israel junction condition shows that the answer is NO. The
coordinate radius of the normal sphere must shrink. Heuristically
this is because the pressure of the DBHF at fixed entropy is larger
than that of the normal region.  On the other hand, if the universe
is future asymptotically dS, then the causal diamond of a maximal
observer is a marginally trapped null surface of fixed area. We can
join it onto the event horizon of a black hole of equal area,
embedded in the $p = \rho$ FRW background, satisfying the Israel
condition. Thus, holographic cosmology {\it predicts} that the
normal region of the universe must asymptote to dS space, with a
c.c. determined by cosmological initial conditions.   Since the DBHF
is infinite, we can choose initial conditions in which it is seeded
with an arbitrary finite number of normal regions, each with its own
c.c.   This is a {\it multiverse} in which the regions with
different physics never talk to each other, but it nonetheless
provides a model in which the c.c. is subject to environmental
selection constraints.

The DBHF defect cosmology evolves to a matter dominated FRW universe.
On the equal area time slices, regions of normal universe grow in volume relative to those regions which behave like the DBHF. Even if the defect was a small fraction of the initial volume, the eventual picture is that of a normal universe seeded with small regions of DBHF, each of which will just become a black hole in the normal universe.  Note however that this picture is self consistent only if the resulting distribution of black hole masses, velocities and positions is close to a homogeneous isotropic fluid.  Otherwise, black holes will collide and grow and the system collapses back to the DBHF. Thus, in this model, the universe undergoes a transition from a DBHF to a dilute, nearly homogeneous and isotropic black hole gas.

Holographic cosmology thus provides an explanation for the homogeneity, isotropy, flatness and low entropy of the initial normal universe, without invoking inflation\footnote{In fact, I believe, with Penrose, that the inflationary explanation of homogeneity, isotropy and low initial entropy is not valid.}  It cannot however explain two salient properties of the Cosmic Microwave Background: the long range, visible universe spanning, correlations between fluctuations, and the fact that these fluctuations enter the horizon oscillating in phase. The latter fact is responsible for the acoustic peaks that are observed in the data.

One must thus postulate that the particle physics of the normal universe contains a field with the properties of an inflaton. Note that the initial conditions imposed by the DBHF explain why {\it the initial inflaton field is homogeneous over the whole region which evolves to our current horizon}. If it were not homogeneous at the time when the universe is dominated by the dilute black hole gas, black holes would again collide and cause collapse to the DBHF.

There are two sources of fluctuation in the inflaton energy density after inflation. The first comes from initial conditions.  Locally, the transition from the dilute black hole gas to an inflationary era comes about when the energy densities of the two become equal. Thus, fluctuations in the initial black hole density, which must be small but need not be zero, are imprinted on the evolution of the inflaton field. In addition, we have the usual quantum fluctuations of the inflaton. One can show that the former lead to a slightly blue tilted or at best exactly scale invariant spectrum\footnote{The spectrum of fluctuations in the black hole density is exactly scale invariant over a finite range of scales, and zero outside that range. The blue tilt comes from evolution of the inflaton field.}, while the latter are famously red-tilted. The data seem to favor quantum fluctuations as the origin of the fluctuations in the CMB, but it is not ruled out that there is some contribution from the initial black hole density fluctuations, on the scales important for structure formation.

Note that the number of e-folds of inflation that are required to explain the CMB data is of order $20$, considerably fewer than one postulates in conventional inflationary cosmology. This alleviates the problem of fine tuning of the inflaton potential.  It is also likely that the scale of inflation is much lower than in conventional models. The universe must go through the DBHF phase, and the dilute black hole gas phase, before inflation begins.

This ends our brief recapitulation of holographic cosmology. We next turn to study the variables of the holographic quantum theory.

\section{SUSY and the holographic screens}

The variables of quantum gravity are related to the classical
concept of the orientation of a pixel on the holographic screen of a
causal diamond. For a finite area causal diamond, the holographic
principle implies a UV cutoff on the number of functions describing
the geometry of the screen.  We can implement this by replacing the
algebra of functions on the screen by a finite dimensional
associative algebra. It turns out that in order to describe
particles, we should choose this to be a non-commutative matrix
algebra. A pixel simply means a single element of a basis of this
algebra\footnote{We will use the phrase {\it localized pixel} for
the more conventional geometrical notion of pixel - a basis of the
algebra satisfying $f_i f_j \approx \delta_{ij}$, in the large area
limit.}. Following the rules of non-commutative geometry, vector
bundles over the screen are then rectangular matrices.

The orientation of pixels on the screen is described by a section of
the spinor bundle over the screen.  Indeed, the screen is a leaf of
a foliation of the null boundary of the causal diamond. That is to
say, there is a null direction penetrating each pixel of the screen.
The Cartan-Penrose equation

$$\bar{\psi} \gamma^{\mu} \psi\  (\gamma_{\mu})^{\alpha}_{\beta}
\psi^{\beta} = 0 ,$$ defines a null vector in space-time and a $d
-2$ plane transverse to it. The components $\bar{\psi} \gamma^{\mu_1
\ldots \mu_k} \psi $, with $k \geq 2$ all lie in this plane. The
local Lorentz and scaling symmetries of the CP equation are gauge
symmetries (the scaling symmetry is broken to a $Z_2$ gauge
symmetry, which we identify with $(- 1)^F$, by the quantization
rules below). In the limit of a large screen, a localized pixel
uniquely determines a null direction up to an ingoing-outgoing
ambiguity. Given that null direction, the
solution of the CP equation is just a null plane spinor,$S_a$ a
spinor under the transverse rotation group. Thus, the screen
orientation variables live in the spinor bundle over the screen.

If we are describing a $d$ dimensional infinite space, with
asymptotic $SO(d - 1)$ symmetry, the most general commutation
relations compatible with the holographic principle, for single
pixel variables is
$$ [S_a , S_b ]_+ = \delta_{ab},$$ where for simplicity we have
chosen a dimension where the transverse spinor representation is
real.  SUSY aficionadas will recognize this as the commutation
relations for the spin degrees of freedom of a $d$ dimensional
massless supermultiplet  - {\it the quantized screen orientations on
a pixel have a space of states identical to the spin states of a
massless superparticle}.  For multiple pixels, we should expect the
independent degrees of freedom to commute, but we can use the $Z_2$
gauge symmetry to write the quantum algebra as
$$ [S_a (m) , S_b (n) ]_+ = \delta_{ab} \delta_{mn},$$ where we
recall that the labels $m,n$ refer to a basis in the spinor bundle
over the fuzzy screen, which is a space of rectangular matrices.

\subsection{Holographic compactification}

In the limit of an infinite, rotationally invariant holographic
screen, these commutation relations lead to the space of states of a
massless super-particle\cite{11D}\cite{bfm}.  However, if the number
of non-compact dimensions is $d < 11$ they do not automatically lead
to super-gravitons\footnote{If $d > 11$ they lead to particle
spectra that are compatible only with a trivial S-matrix.}. For $d =
4$, in the fully $SO(3)$ invariant formulation of \cite{bfm} they
lead only to massless chiral multiplets.  String theory suggests an
obvious solution to this problem, namely the introduction of compact
dimensions. For compactifications to $d = 4$ from $11$ dimensions
(or $10$ with Type II spinors), the spinor bundle is described by
variables $(\Psi^I)_i^A , (\Psi^{\dagger J})_B^j $, where $I,J$ are
$8$ component spinor indices.  The $N \times N + 1$ and $N+1 \times
N$ matrices $\Psi$ and $\Psi^{\dagger} $ are the two spinor bundles
over the fuzzy two-sphere.

The anti-commutation relations are

$$ [(\Psi^I)_i^A , (\Psi^{\dagger J})_B^j ]_+ = \delta_i^j
\delta^A_B M^{IJ} , $$ and the geometry of the compactification is
encoded in the super-algebra formed by the bosonic generators
$M^{IJ}$ and the pixel orientation variables $\Psi$ and
$\Psi^{\dagger}$.  Indeed $M^{IJ}$ are in the bi-spinor over the
compact space, which is a sum of $p$-form charges. From experience
in string theory, we expect to write each $p$-form charge as a sum
over wrapped brane charges associated with cycles.  These will
generically form an abelian bosonic algebra, but on singular
manifolds, when some cycles shrink to zero size it can become
non-abelian.  For $N \rightarrow\infty$, we expect to be able to
achieve $4$ dimensional super-Poincare invariant
compactifications\footnote{From the holographic point of view, AdS
space-times are quite different from either asymptotically flat, or
dS space. In the latter two cases, causal diamonds with finite
proper time separation between past and future tips, all have finite
holographic area.  In AdS space, the area becomes infinite in finite
proper time, and this is the reason that the conformal boundary of
space-time is time-like and its quantum dynamics is described by a
quantum field theory. In dS or asymptotically flat space, the
dynamical object for local observers is the scattering
matrix\cite{holo1} (which has small ambiguities in the dS case),
mapping the past and future boundaries of the causal diamond onto
each other.}. Partial SUSY breaking will be encoded in the fact that
the bosonic generators $M^{IJ}$ do not commute with all of the
$\Psi^I$. Holographic compactifications can be classified by finding
all such algebras whose $N \rightarrow \infty$ spectrum contains the
supergravity multiplet for some super-Poincare algebra in four
dimensions.

One of the salient features of this approach to compactification is
that it is manifestly invariant under string-dualities.  A well
known feature of such dualities is that the SUSY algebra is
invariant, and only its interpretation in terms of branes wrapped on
cycles of a large smooth manifold, Kaluza-Klein momenta, or purely
internal symmetries, changes from one duality frame to another.

The actual geometry of the internal manifold is determined by the
explicit matrix representation of the pixel super-algebra, which
must, by the holographic principle, be finite dimensional.  This
means that arbitrarily large internal manifolds can only arise by
some sort of correlated limit as $N \rightarrow\infty$.  In
particular, for dS space, where the total entropy is finite, the
size of the internal manifold is bounded. We will discuss this in
more detail in the next subsection.  More generally, the
discreteness of finite dimensional representation theory assures us
that {\it all} moduli of the theory are frozen in dS space.
Continuous moduli spaces are artifacts of the infinite entropy
asymptotically flat limit.

Implicitly, we have assumed that the algebra of functions on the
internal manifold is finite dimensional and it is likely to be
non-commutative.  In the large volume limit, a remnant of
non-commutativity would show up as a Poisson structure on the
manifold. Many of the standard compactification manifolds of string
theory have such a structure.  Many of them are compact Kahler
manifolds, and one could find sequences of matrix algebras
approximating $C^{\infty} (M)$, by the methods of
geometric/deformation quantization.   The seven manifolds that
appear in compactification of M-theory to four dimensions also have
natural Poisson structures.  For example, Horava-Witten
compactification is a Calabi-Yau bundle over an interval, and there
is a natural Poisson structure induced by the Kahler form of the
Calabi-Yau manifolds. Manifolds of $G_2$ holonomy are less well
understood, but the construction of non-compact manifolds of this
type can usually be understood as the lift of D6 branes in Type IIA
string theory on a non-compact Calabi-Yau 3-fold, so the 7 manifold
is a circle bundle over the Calabi-Yau 3-fold, and there is again a
natural Poisson structure.

\subsection{Matrices and particles}

Suppose that we have found a sequence of matrix algebras ${\cal
A}_n$, of $n \times n$ matrices, which converges, in an appropriate
sense, to the algebra ${\cal F}$ of functions (with some continuity
or smoothness property) on the holographic screen at infinity for a
space-time of the form $M^{1, d - 1} \times X$. We can consider the
block matrices
$$ \begin{pmatrix} M_{k_1} & \ldots & 0 \\ & \ldots & \\ \vdots & &
\vdots \\  & \ldots &  \\ 0 & \ldots & M_{k_p} \end{pmatrix} ,$$
where $M_k$ is a $k\times k$ matrix in ${\cal A}_k$ .  We can view
this as a subalgebra of all $N \times N$ matrices with $N = \sum
k_i$.

Taking the limit where all $k_i$ go to infinity, at fixed ratio, we
can obtain not just single super-particle states, but an entire Fock
space.  In addition, the (in the limit continuous) block sizes
$k_i$, supply the magnitude of the null momenta of these particles,
which was missing in our kinematic description above. The
permutation gauge symmetry of particle statistics is a natural
consequence of the unitary equivalence of different bases in the
full $\sum k_i \times \sum k_i$ space of matrices. Combining this
with our $(- 1)^F$ gauge symmetry, we get the correct connection
between spin and statistics.

Technical details of the limit, which lead to Lorentz invariance,
involve a famous infinite dimensional operator algebra, discovered
by Murray and von Neumann.  These details can be found in
\cite{11D}.  This construction raises the question of what to think
about the rest of the matrix degrees of freedom, apart from the
diagonal blocks.  We will see that the quantum theory of dS space
has an answer to this question. They represent particles propagating
in horizon volumes causally disconnected from ``our own''.

\section{Quantum interpretation of semi-classical properties of dS space}

One should take seriously those aspects of
quantum gravity in dS space that can be reliably described in the
semi-classical approximation.  Among these are

\begin{itemize}

\item The dS vacuum is actually a density matrix with high entropy
equal to $\pi (R M_P)^2$, where $R$ is the dS radius, related to the
c.c. by $R^{-1} = \frac{3 m_P}{\Lambda^{1/2}}$. According to the
strong form of the holographic principle advocated above, this
entropy is the logarithm of the dimension of the Hilbert space
describing stable dS space.

\item There is an operator we will call $P_0$, which converges to
the Minkowski Hamiltonian in a particular Lorentz frame (that of a
given static observer in dS space) in the limit $R M_P \rightarrow
\infty$. The mass parameter in dS black hole solutions (see below)
is interpreted as the eigenvalue of this operator. With respect to
this operator, the density matrix is a thermal state, with a {\it
unique} temperature, $ T = \frac{1}{2\pi R}$. The instanton
calculation of \cite{gpbh} shows that the probability of nucleating
black holes of small mass is also thermal, with the same
temperature.

\item Although one often claims that the dS group converges to the
Poincare group through Wigner-Inonue contraction, a proper GR
interpretation of generators in terms of their action on the
boundary of a causal diamond, contradicts this. Near the future
cosmological horizon ($ v \rightarrow 0$ below) the dS metric looks
like
$$ ds^2 = R^2 ( - dudv + d\Omega^2 ),$$ and the generator of static
time translation is $ \propto (u\partial_u - v\partial_v)$, while
the Minkowski metric near future null infinity is
$$ ds^2 = \frac{ - du dv + d\Omega^2 }{v^2} ,$$ and the time
translation is $\propto \partial_u $. This suggests that the quantum
theory of dS space has another Hamiltonian, in addition to $P_0$.

\item Instanton transitions between two dS minima of an effective
potential always\footnote{As long as the potential at the top of the
barrier is not too flat. There are reasons to believe that this
degree of flatness is not consistent with quantum
gravity\cite{bdfgahmv}.} exist, and satisfy a principle of detailed
balance $$P_{1\rightarrow 2} = P_{2\rightarrow 1} e^{- (\Delta
S)_{12}},$$ appropriate for a system at infinite temperature (free
energy equals entropy times temperature).  For transitions to
negative c.c. Big Crunches, the instanton calculation gives a
probability $ \sim e^{- \Delta S}$, if one interprets the entropy of
the Crunch state as one quarter of the area of the maximal causal
diamond of an observer in the crunching region, but only for
potentials which are {\it above the great divide}\cite{abj}. If, for
any given dS minimum of a potential, one shifts the energy down to
zero, one can ask whether the resulting Minkowski solution has a
positive energy theorem. The space of potentials that, after such a
shift, have a static domain wall between the Minkowski solution and
an AdS solution in the negative minimum, is co-dimension one in the
space of all potentials.  On one side of this great divide there is
a positive energy theorem, and a dS solution gotten by shifting the
potential upwards will have the $e^{ - \Delta S}$ behavior. We
interpret this as saying that only potentials above the divide can
appear in the low energy effective field theory of stable dS space.

\item Schwarzschild black hole solutions\footnote{The conclusions we
will draw here also hold for charged black holes. We have not yet
modeled spinning black holes in the quantum theory.} in dS space
have the form
$$ds^2 = - dt^2 f(r) + \frac{dr^2}{f(r)} + r^2 d\Omega^2 ,$$ where
$f(r) = 1 - \frac{2M}{M_P^2 r} - \frac{r^2}{R^2} .$  These have two
horizons with
$$ R^2 = R_+^2 + R_-^2 + R_+ R_- ,   \ \ \ 2M = M_P^2 \frac{R_+ R_-
(R_+ + R_-)}{R^2} .$$ Note that the sum of the areas of the two
horizons is less than the vacuum area and decreases with increasing
$M$, reaching a minimum at the maximum mass black hole $M_N =
\frac{M_P^2 R}{3\sqrt{3}}$.

\item The maximum entropy in a dS horizon volume, which can be described by
quantum field theory, without invoking black holes of order the
horizon scale, is $ \sim (RM_P)^{3/2} $.

\end{itemize}

\subsection{The quantum theory of stable dS space}

We now sketch a quantum theory of dS space based on the pixel
variables introduced above and consistent with all of this
semi-classical ``data''.  The variables consist of $N(N+1)$ copies
of the single pixel algebra, so the entropy, to be identified with
$\pi (RM_P)^2$ in the large $N$ limit is  $N^2 {\rm ln} D_{\cal P}$,
where $D_{\cal P}$ is the dimension of representation of the single pixel algebra.

On the other hand, by standard Kaluza-Klein arguments the entropy
per pixel is given by
$$ {\rm ln} D_{\cal P} = (\frac{m_P}{m_d})^2 ,$$ where $m_d$ is the
higher dimensional Planck mass.  If, following
Witten\cite{wituniplanck}, we identify this with the scale of
coupling unification, then  $${\rm ln} D_{\cal P} \sim 10^4 .$$

According to our general discussion, particle states of the system
will arise by restricting attention to block diagonal matrices, of
the form

$$\begin{pmatrix} \Psi_{k_1} & \ldots & 0 \\ & \ldots & \\ \vdots & &
\vdots \\  & \ldots &  \\ 0 & \ldots & \Psi_{k_p}  \end{pmatrix} .$$
If we want to maximize the entropy in such particle states, we
distribute it evenly among the different particles.  We would like
to have a lot of particles, but each one should have a lot of
angular momentum states, so that we can localize it on the sphere
and justify the name particle.  The compromise is to have $N^{1/2}$
blocks of size $N^{1/2} \times N^{1/2}$.  Since $N \propto (RM_P)$
this gives an entropy scaling for particles in a single horizon
volume identical to that derived in our semi-classical argument.

In this context there is an obvious interpretation of the other
degrees of freedom in the matrix variables. Indeed they can be
organized into $N^{1/2}$ bands, each consisting of a collection of
blocks identical to the ones we have associated with particles in a
single horizon volume.  The semi-classical theory of dS space seems,
at very early and late times, to accommodate an infinite number of
copies of the particle degrees of freedom in a single horizon
volume.  Here we see that there is an organization of the Hilbert
space into $N^{1/2}$ such copies.  Since $N \sim M_P R$, this agrees
with the semi-classical limit in which $M_P \rightarrow \infty$ with
the dS radius fixed in centimeters. We will see later that this
decomposition breaks down when we begin to consider black hole
states of the system.

In order to explain the gross properties of semi-classical dS space,
we introduce a Hamiltonian $H$, which may be viewed as the evolution
operator with respect to the static time coordinate. As noted this
should be different from, and should not commute with, the
Hamiltonian $P_0$ which is the sum of localized particle
Hamiltonians in each horizon volume\footnote{This statement should
be understood in the sense of scattering theory.  All localized
states in dS space decay to the vacuum by emission of particles, and
the particles are non-interacting when they are far apart, so one
can calculate the energy just by summing up the kinetic energies of
particles.}. $H$ is the Hamiltonian whose dynamics is probed by the
instanton computations we have described above.

We model it as a Hamiltonian with a random spectrum, obeying the
bound $|| H || \leq c T$.  By random spectrum, I mean that time
evolution under $H$ obeys the rules of statistical mechanics: for
any initial state, the time dependent expectation values of a large
class of observables will evolve rapidly to thermal equilibrium,
with a temperature determined by the initial expectation value of
$H$\cite{deutsch}.  For a random choice of initial state, the
temperature will be infinite, with probability $1$. If we interpret
the dS vacuum state, following the holographic principle, as the
infinite temperature ensemble, then we can understand the gross
features of the probabilities of Coleman-DeLucia tunneling events in
dS space, assuming the low energy effective theory is above the
divide.

To explain the fact that same density matrix is thermal, with a
finite temperature, for the Hamiltonian $P_0$ we need only postulate
a relationship between the eigenvalues of $P_0$ and the degeneracies
of the corresponding eigenspaces.  This corresponds to a formula
$$P_0 = \sum E^I e_I ,$$ where $e_I$ are commuting orthogonal
projectors satisfying
$${\rm Tr}\ e_I \sim e^{\pi (RM_P)^2 - \frac{p_0^I}{T}}, $$
where the $\sim$ sign means equality to leading order in
$\frac{p_0^I}{m_N}.$ $m_N = \frac{m_P^2 R}{3\sqrt{3}}$, is the mass
of the maximal, Nariai, black hole in dS space.  $P_0$ is an
operator defined only for states localized in a single horizon
volume.  That is, we view the full Hilbert space of the stable dS
theory as a tensor product of a space of states localized in a
single horizon value, with its complementary tensor factor.  $P_0$
acts only in the localized factor {\it i.e.} it acts as the unit
operator in the complementary factor.   We know that the maximal
entropy density matrix describable by field theory in the single
horizon has localized entropy of order $(RM_P)^{3/2}$.  For the
maximal size black hole, the entropy (counting both black hole and
cosmological horizons) approaches $\frac{2}{3} (RM_P)^2$.  This
indicates that all localized states have entropy much smaller than
that of the dS vacuum and that entropy decreases with increasing
$P_0$.  The maximal eigenvalue of $P_0$ is the Nariai black hole
mass. The formula above incorporates both the predictions of field
theory, of a thermal distribution at the dS temperature for particle
states, and the instanton calculation of \cite{gpbh}, which
indicates a thermal probability at the same temperature for black
hole nucleation.

There is some semi-classical evidence for this connection between the energy and entropy of eigenstates of $P_0$.  For eigenvalues much larger than the Planck scale, but much smaller than the Nariai mass, the black hole entropy formula in dS space gives
$$S (M) = S_{dS} - \frac{M}{2\pi R} , $$ which is just the energy/entropy connection we have postulated above.

To connect the evolution under the Hamiltonian $H$ (which produces
the dS ground state density matrix $\rho = e^{-S} 1$, ) to that of
$P_0$, we have to postulate commutation relations between these two
operators.  These are motivated by the geometrical picture of the
action of the static dS and static Minkowski Killing vectors on the
boundaries of the corresponding causal diamonds (Item 3 in our list
of the properties of dS space)
$$[(u\partial_u - v\partial_v), \partial u ] = \partial_u .$$
In the quantum theory we postulate
$$ [H,P_0 ] =   M_P^2 g(\frac{P_0}{M_P^2 R}), $$ where $g(x)$ is a
bounded function, linear in $x$ for small $x$, and with vanishing
trace.  This coincides with a properly normalized version of the
Killing vector commutation relation in the subspace with $P_0$
eigenvalues $\ll$ the Nariai mass.

The spectrum of $H$ is highly degenerate, with level spacings as
small as $\frac{1}{R}e^{ - \pi (R M_P)^2}$. The maximum spacing is
of order $1/R$, which gives evolution on a time scale roughly the
current age of the universe. $P_0$ breaks the degeneracy and evolves
the system on much shorter time scales.  The commutation relations
guarantee that the low lying eigenstates of $P_0$ will be relatively
stable under $H$ evolution. Other approximate quantum numbers could
give rise to stability on even longer time scales.   For example,
although charged particles centered at the origin of the static
coordinate system are not stable, the mechanism of decay is
nucleation of a particle of opposite charge, which has probability
$e^{- 2\pi R m_c}$, where $m_c$ is the mass of the lightest charged
particle. One would conjecture that to reproduce this behavior all
that is necessary is that the Poincare invariant
$(RM_P)\rightarrow\infty$ limit have a massless $U(1)$ gauge boson
in its spectrum.

\subsection{Black holes and pixels}

We will now show that the same variables that describe particle excitations of dS space can describe the properties of Schwarzschild-de Sitter black holes.  Recall the equations relating the two horizon radii

$$ R^2 = R_+^2 + R_-^2 + R_+ R_- ,   \ \ \ 2M = M_P^2 \frac{R_+ R_-
(R_+ + R_-)}{R^2} .$$
Recall also that

$$ \pi (RM_P)^2 = {\rm ln} D_{\cal P} N^2 $$ and define
$$ \pi (R_{\pm} M_P)^2 = {\rm ln} D_{\cal P} N_{\pm}^2 $$
Choose $N_-$ to be an integer and $N_+$ to be the greatest integer in the solution of
$$ N^2 = N_+^2 + N_-^2 + N_+ N_- $$

We define the ensemble of black hole states to be the states satisfying
$$ \psi_i^A | BH \rangle = 0$$ for $1 \leq i \leq N_-$, $1 \leq A \leq N_+$.
The entropy of this ensemble is, for large $N$ and $N_-$ the same as that of the corresponding black hole.  Note that we had to choose a basis for the matrices in order to define this ensemble.  As with particle states, this corresponds to choosing the horizon volume in which the black hole sits.
However, for general large $N_-$, of order $N$, there will be no multiple black hole states.

The fermion number operator has a large expectation value and small statistical fluctuations in this ensemble.  Using the above formula for the black hole mass one can write a function of the number operator whose statistical expectation value $ = M$ in this ensemble, thus realizing the black hole spectrum in this quantum mechanical model.  The precise expression can be found in \cite{bfm}, and is not particularly illuminating or edifying.

What we see from these formulae is that the same variables can describe both black holes and particles, and that states with too many or too energetic particles will inevitably look like black holes.

\section{Constraints on the low energy effective action}

In the previous sections, I've outlined a quantum theory of stable
dS space and indicated how to make it compatible with all of the
semi-classical data we have about the nature of quantum dS space.
Now we will proceed to discuss the constraints on low energy
particle physics which follow from this picture.  The first of these
is that we should view the c.c. as a tunable input parameter, rather
than a calculable quantity. Indeed, one might hope that it is the
{\it only} tunable parameter in the model.  There is no real
argument for this extreme position, but there certainly are
arguments that stable dS models are a lot fewer and further between
than other classes of string models, like supersymmetric AdS models
and models of asymptotically flat space.

The mere fact that the c.c. is tunable means that for small c.c. we
should be able to model the dynamics of localizable states by a low
energy effective Lagrangian density.  As usual in string theory we
view the construction of this Lagrangian as an act of {\it
theoretical phenomenology}. The underlying theory prescribes some
properties (which in perturbative string theory, but only in
perturbative string theory, correspond to a systematic calculation
of all observables in an expansion in some dimensionless parameter),
which must be reproduced by the Lagrangian. In our case, these
properties include the value of the c.c. and the fact that the
limiting Lagrangian when $R M_P \rightarrow \infty$ must describe a
consistent coupling to some super-Poincare gravity multiplet. As a
consequence, the model must have a dS solution, with a c.c. that can
be tuned to zero.  As far as I know, this implies that we have $3 +
1$ non-compact dimensions\footnote{It is certain that there are no
super-Poincare invariant gravitational scattering matrices in $1 +
1$ dimensions, and likely that the same is true in $2 + 1$
dimensions. Higher dimensional supergravities coming from unitary
theories, and higher ${\cal N}$ supergravity in four dimensions do
not appear to have dS solutions.} with ${\cal N} = 1$ SUSY.
Furthermore, since SUSY is a global limit of a gauge symmetry, its
breaking must occur through the super-Higgs mechanism.

Generic SUGRA Lagrangians and generic supersymmetric stationary
points, have negative rather than vanishing c.c.  A natural way to
account for the vanishing of the c.c. in the supersymmetric limit is
to postulate that the low energy Lagrangian in this limit has a
discrete R symmetry $Z_k$ with $k \geq 3$. This ensures the
vanishing of the superpotential at supersymmetric, R-symmetric
stationary points.  The terms that must be added to the Lagrangian
when $\Lambda > 0$ explicitly break this R-symmetry, since they must
in particular include a constant $W_0$ which will tune the c.c. to
$\Lambda$ when the scale of SUSY breaking takes on its correct value
(which we'll discuss in a moment). When these terms are discarded
the Lagrangian must preserve SUSY, and their addition is what
triggers SUSY breaking.

The underlying model also suggests a scale for the gravitino mass.
We have seen that single particle excitations of the model arise
from fermionic matrices of size $K$, where maximum entropy
multi-particle states have\footnote{Here we have used the relation
between the total entropy and the dS radius in four dimensional
Planck units, as well as the KK estimate of the number of degrees of
freedom per four dimensional pixel. $m_D$ is the higher dimensional
Planck scale.} $K \sim N^{1/2} \sim \pi^{1/4} (\frac{m_D}{m_P}
RM_P)^{1/2}$. There is a precise action of the rotation group on
these states, but the maximum spherical harmonic in their wave
functions has $L \sim N \sim 10^{60}$. Lorentz boosts act on the two
sphere at infinity as the coset of rotations in its $SO(1,3)$
conformal group. The breaking of Lorentz invariance due to the
restriction to only $10^{60}$ spherical harmonics is too small to be
measurable by current experiments.

On the other hand, it is reasonable to suppose that the breaking of
SUSY is governed by $N^{- 1/2}$, the natural expansion parameter of
this system.  That is, the commutator of the SUSY generators with
the Hamiltonian will be of order $N^{- 1/2}$ in Planck units. If we
choose $m_D$ to be the unification scale, $\sim 2 \times 10^{16}$
GeV, we get the estimate for the gravitino mass\footnote{In the
super-Higgs phase, the super-partner of any particle state is that
state plus a longitudinal gravitino.}
$$m_{3/2} = 10 K \Lambda^{1/4} = K (10)^{1/4} 10^{-2} {\rm eV}.$$

A similar estimate is obtained from the following hand waving
argument. Our underlying model implies that the gravitino mass
vanishes in the limit $RM_P \rightarrow\infty$.   Assume that the
gravitino is the lightest R-charged particle in the theory. Then,
the terms in the effective Lagrangian that lead to the breaking of
SUSY must arise from interactions between the gravitino and states
localized on the dS horizon.  Those states cannot be modeled
properly  by effective field theory. A better model is Landau levels
on a two sphere, with an entropy given by the dS entropy. I use this
analogy because there is a basis of Landau level states which is
localized on the sphere, and we can talk about the entropy of states
in a given area $A$. Assuming that the gravitino mass vanishes like
a power of $RM_P$ in the limit that this parameter goes to infinity,
we can write a term in the effective Lagrangian coming from a
gravitino going out to infinity as (to leading exponential order)

$$\delta {\cal L} \sim e^{ - 2 m_{3/2} R} \sum |\langle 3/2 | V | n
\rangle |^2 . $$ $V$ is a `` vertex operator" coupling the gravitino
to the degrees of freedom on the horizon. Note that there are no
large energy denominators in this formula. The horizon dynamics all
takes place over a much longer time scale than the duration of the
gravitino's sojourn on the horizon. Indeed, since the gravitino is
massive and the horizon is null, it can only be viewed as
propagating near the horizon for a proper time of order
$m_{3/2}^{-1}$. Free quantum particles perform random walks and the
natural step size for this walk on the horizon is the Planck scale.
Thus, one expects the horizon area covered by the gravitino to be
$\frac{1}{m_{3/2} M_P}$, and the number of states for which the
matrix element is non-zero to be of order $e^{\frac{b
M_P}{m_{3/2}}}$, where $b$ is one of those proverbial constants of
order one.

In order for the expression $e^{- 2 m_{3/2} R +
b\frac{M_P}{m_{3/2}}}$ to be neither exponentially small or large in
the large $R$ limit, $m_{3/2}$ must converge to $ \frac{b}{2} M_P
(RM_P)^{- 1/2}$, which is the $N^{- 1/2}$ scaling we postulated
above. Note that the conventional ``no fine tuning" prediction of
SUGRA is $m_{3/2} \sim R^{-1}$. This {\it is} the right scaling in
AdS space, but would lead to an inconsistent, exponentially large
correction from interactions with the boundary.  Of course, this
scaling is already inconsistent with our underlying model.  A
gravitino at rest with such a mass would, according to our formalism
have only one or a few spherical harmonics in its wave function on
the sphere, and would never behave as a localized particle.

\subsection{Particle physics}

The low SUSY breaking scale implied by the above arguments rules out
almost all extant models of SUSY breaking.  In particular,
gravitationally coupled hidden sector models are ruled out, as well
as all gauge mediated models with weakly interacting messengers. The
same is true for anomaly mediation, gaugino mediation, semi-direct
mediation, {\it etc.}. In all of these models, the observed lower
bounds on super-partner masses are inconsistent with our estimate
for the gravitino mass.   Extra factors of $10-10^3$ in the latter
estimate {\it could} make CSB compatible with some of the low energy
mechanisms, but one must also be careful to avoid the window of
gravitino masses ruled out by cosmology, which is something like
$20$ eV to $1$ GeV.   This rules out $F$ terms between $2\times 10^5
{\rm GeV}  \sqrt{F} < 10^9 {\rm GeV}$.  The nominal value of
$\sqrt{F} $ in CSB is $\sqrt{F} \sim 10^4$ GeV. Even $2 \times 10^5$
GeV is a low scale for most conventional gauge mediated models.

Before starting to put in detailed phenomenological constraints, let
me note that the simplest model consistent with the general ideas of
CSB has a low energy sector consisting of a single chiral
superfield, $G$, with zero charge under the fundamental discrete
R-symmetry. The R-breaking terms consist of a general
superpotential, which is a function of $G/m_P$ with the linear term
adjusted to give the CSB formula for the gravitino mass and the
constant tuned to set the c.c. equal to its underlying value
$\Lambda$.  Another important constraint is that the decay
probability to any lower minimum of the potential (there is likely
to be a supersymmetric minimum with negative c.c.) vanish as $e^{-
\pi (RM_P)^2} $ when $\Lambda$ is taken to zero.  In the language of
\cite{abj}, the potential must be {\it above the Great Divide}. This
involves order $1$ adjustments in the dimensionless expansion
coefficients in $W(G/m_P )$.

As soon as we insist on standard model gauge symmetries, things
become complicated.  First note that pure non-abelian gauge groups
are not allowed, we must have matter fields in fairly large
representations (not necessarily irreducible).  Non-abelian SUSY
gauge theories with small representations tend to break all R
symmetries and are not consistent with a super-Poincare invariant
$\Lambda = 0 $ model.  For example, the standard model gauge group
with one generation is not consistent with CSB at $\Lambda = 0$. QCD
with two flavors would generate an R-violating superpotential, and
does not give a super-Poincare invariant space-time.  We will see
that phenomenological considerations put stringent upper and lower
bounds on the number of low energy fields charged under the standard
model.

Indeed, having introduced the standard model, we must enforce the
experimental lower bounds on the chargino and gluino masses. In
terms of the Goldstino superfield $G$, these will come from a term
of the form

$$\beta_i \frac{G}{M} (W_{\alpha}^{i})^2 ,$$ with the $\beta_i$
dimensionless constants of order $1$. Recalling that,according to
the CSB relation between the gravitino mass and the c.c.,  $F_G/ m_P
= m_{3/2} \sim 1.75 \times 10^{-11}$ GeV, the bounds on the chargino
mass imply that
$$M \leq \frac{\beta_2}{1.7} {\rm TeV}, $$
where we have used the strongest possible chargino bound ($> 160 $
GeV from the Tevatron)\footnote{This bound is valid in a class of
models, but may not be general.  The model independent bound is
$\sim 105$.}.  There is also an experimental lower bound of order
$300$ GeV on the gluino mass. In the model we eventually settle on,
there is another parameter, which can give a variety of values for
the gluino mass, below the vanilla gauge mediated prediction
$$m_{1/2}^{(3)} = \frac{\beta_3 \alpha_3}{\beta_2 \alpha_2}
m_{1/2}^{(2)}. $$ We note that the parameters $\beta_i$ lend an
additional uncertainty to these estimates.

We learn two things from this calculation: there must be new states,
charged under the standard model with mass of scale $\sim M$, and
$M$ is quite low. Note that we cannot push the chargino mass much
higher than its experimental lower bound, because that would bring
down the scale $M$, and contradict the experimental absence of these
new states. Thus, models based on CSB predict light chargino states,
close to the current experimental bounds. The particular model we
choose, the Pyramid Scheme, also predicts a gluino mass somewhere
below the ``vanilla gauge mediation'' value quoted above.

Note also that the gravitino is very light, and consequently
cosmologically safe.  It is however, the LSP and the NLSP (which is likely to be either the bino or a right handed
slepton) is coupled to it strongly. Conventional SUSY dark matter is
not obtained in this model. We will discuss dark matter in the
context of the Pyramid Scheme below.

When combined with the requirement of coupling unification, the low
scale $M$ for additional matter with standard model quantum numbers,
puts strong constraints on the nature of the hidden sector whose
dynamics becomes strong at the scale $M$. In order to preserve one
loop unification predictions in a natural manner, the hidden sector
matter must be in complete multiplets of the unified group. On the
other hand, if the hidden sector group is large, this extra matter
will produce Landau poles below the unification scale, destroying
the prediction of unification.

To this constraint, we must add the fact that the strong dynamics of
many hidden sector models is phenomenologically unacceptable, or
violates the principles of CSB.  The latter is true for example if
the hidden sector theory preserves SUSY, but not an R symmetry. The
phenomenological problems are connected to the existence of light
pseudo-Nambu-Goldstone bosons (PNGB) in the hidden sector.

While an exhaustive search has not been carried out, the only model
that has been found, which is compatible with unification in $SU(5)$
or some larger simple group, is the Pentagon model of
\cite{pentagon}. However, this model has Landau poles below the
unification scale, and contains a PNGB whose interactions with
electrons lead red giant stars to cool too rapidly\cite{tbhh}. In
addition, discussions of this model have concentrated on vacuum
states that are meta-stable in the field theory approximation,
neglecting gravitational effects. We will argue below that no such
model can truly represent the physics of CSB.

Before turning to that discussion we introduce the most promising
model of CSB that has so far been found, the Pyramid
Scheme\cite{pyrma12}. The basic idea is to replace unification in a
simple group by Glashow's {\it trinification}, the semi-direct
product of $SU_1 (3) \times SU_2 (3) \times SU_3 (3) \ltimes Z_3$,
where the $Z_3$ permutes the three copies of $SU(3)$, cyclically.
The indices on the different $SU(3)$ groups (almost) indicate which
subgroup of the standard model gauge group is embedded in which
unified $SU(3)$ . The exception is the $U(1)$ of weak hypercharge,
which is a linear combination of the generator in $SU_2 (3)$ that
commutes with the standard model $SU(2)$, and a generator in $SU_1
(3)$. At the unification scale, the matter content of the theory
consists of 3 copies of the equilateral triangular quiver consisting
of $(3,\bar{3},1) + $ cyclic permutations. This can be thought of as
the $27$ of $E_6$, decomposed under the trinification subgroup.
Another, probably more interesting way of looking at it, which even
gets the number of generations correct, is to view this as $3$ $D3$
branes at a $Z_3$ orbifold singularity in Calabi-Yau
compactification of Type IIB string theory. We must arrange the GUT scale
model so that below the unification
scale, only the $15$ chiral fields of an MSSM generation survive. This can probably be achieved with Wilson lines in the compact dimensions, but a complete stringy construction of the Pyramid Scheme is not yet available.
The standard model Higgs fields can arise in a variety of ways.

Now we can introduce a hidden sector gauge group $G$, which extends
the quiver to a three sided pyramid.  The fields that connect the
apex of the pyramid to its base are in the
$$(R, \bar{3}, 1, 1) + (\bar{R}, 3, 1, 1) + {\rm cyclic
permutations}. $$ If the representation $R$ of $G$ is small enough,
we will have one loop unification with no Landau poles. We will make
the choice of the group $G$ and representation $R$ in order to
ensure that the low energy dynamics of the model reproduces the
behavior expected from a quantum theory of dS space with a finite
number of states. We pause to review that behavior in the next
subsection.

\subsection{SUSY breaking and tunneling}

We have already argued that the low energy effective Lagrangian must
be $4$ dimensional ${\cal N} = 1$ SUGRA, with a dS minimum.  It must
contain parameters enabling us to tune both the SUSY breaking scale
and the c.c. to zero, with the relation
$$ m_{3/2} = 10 K \Lambda^{1/4} .$$  In the limit of vanishing c.c.,
it must have a super-Poincare invariant solution with a discrete
complex R symmetry, or perhaps a compact moduli space of such
solutions\footnote{In all known string theory motivated models of
SUSY breaking - SUSY is asymptotically restored in non-compact
directions of moduli space, and the Coleman De Lucia tunneling
amplitudes from the dS minimum to the non-compact region, do not
obey the requirement that the dS minimum represents a stable system
with a finite number of states.  We review this requirement below.}.

A further restriction comes from considering Coleman-De Lucia
tunneling from the dS minimum to other, lower minima of the
potential. In particular, one must generally expect a SUGRA
Lagrangian with a dS minimum, to have other solutions where all SUSY
order parameters vanish. Generally, there will be CDL tunneling from
the vicinity of the dS minimum to the vicinity of the supersymmetric
one. If, as is typical, the supersymmetric minimum has negative
vacuum energy, classical evolution after tunneling leads to a Big
Crunch space-time. The covariant entropy bound then implies that no
observer in the post-tunneling space-time can see more than a
finite, usually small, amount of entropy.

Such a transition makes sense in a finite theory of stable dS space,
if the tunneling probability is of order $$ P \sim e^{ - \pi
(RM_P)^2} $$ as $RM_P \rightarrow \infty$. This is the probability
one expects for a Heisenberg/Poincare recurrence in a finite system
with a number of states equal to the exponential of the dS entropy.
It is a catastrophe analogous to all of the air collecting in the
corner of a room, rather than a true instability.  In \cite{abj} my
collaborators and I did an analysis of the conditions in low energy
effective field theory, which guarantee this kind of behavior. The
space of all possible potentials, modulo an additive constant is
divided in two, separated by a co-dimension one ``Great Divide''. A
precise definition of the divide is obtained by adding a constant to
the potential, to bring the dS minimum in question down to zero
potential energy.

The Great Divide is the submanifold of the space of potentials, on
which there exists a static domain wall solution, connecting the
zero energy minimum to the negative minimum, which has the highest
CDL transition probability when $\Lambda \neq 0$. On one side of the
Divide, the zero energy minimum has a positive energy theorem and no
CDL instability.  On the other side, it decays.  On the side with
the positive energy theorem, we can now restore the positive c.c.
and show that the CDL probability for dS space to decay is that of a
recurrence in a finite system.

It is important to emphasize that the Divide is defined by the
particular one parameter deformation of the potential by an additive
constant.  In particular, any system for which SUSY is restored as
the c.c. is tuned to zero will have a tunneling probability that
behaves like
$$e^{ - \pi (RM)^2} ,$$ as $RM_P \rightarrow \infty$. However, if $M
\ll M_P $ then this is much larger than the probability of a
recurrence.  In particular, this criterion rules out all models of
SUSY breaking in which the dS minimum is meta-stable in the $M_P
\rightarrow \infty$, like the models of \cite{iss}.  In such models,
the potential has the form $$\mu^4 v(\phi / M ),$$ where $\mu$ and
$M$ are $\ll m_P$.  In that case one can show \cite{abj} that the
CDL instanton is essentially the false vacuum dS sphere, with a
small polar cap cut out, in which the non-gravitational
\cite{sidney} instanton is inserted. The action difference, which
defines the transition probability, is of order $(\frac{M}{\mu})^4
$.  The probability indeed vanishes exponentially in the
supersymmetric limit $\mu\rightarrow 0$, but it is {\it much} larger
than that of a recurrence in a finite system with total entropy of
order $(\frac{m_P}{\mu})^4 $.

If we now combine the general features of CSB with an old result of
Nelson and Seiberg\cite{ns}, we come to a rather interesting
conclusion. We have argued that, when gravity is neglected, a system
consistent with CSB has a low gravitino mass and must break SUSY
spontaneously, once we add the appropriate R-violating terms to the
Lagrangian.  Nelson and Seiberg argued that, in such an R-violating
system, spontaneous SUSY breaking was only possible for non-generic
Lagrangians.  The requirement that a Lagrangian be generic, {\it
i.e.} that it contain all terms compatible with symmetries, with
values determined roughly by dimensional analysis, is a consequence
of the renormalization group. Experience shows that this is what one
obtains when integrating out degrees of freedom, even when those are
just the higher momentum modes of the fields in the low energy
theory. It is even true in conventional string theory.

However, in the context of CSB the R violating terms in the
Lagrangian have a very different origin. In our hand waving
argument, they come from Feynman graphs where a single gravitino
interacts with non-field theoretic degrees of freedom on the
horizon. It seems plausible that RG genericity might not hold for
these terms.  Indeed, if one simply insists that the effective
Lagrangian reproduce the basic property of the underlying quantum
theory of dS space, {\it viz.} that it be a finite system with a
stable high entropy density matrix describing the dS ground state,
our argument two paragraphs above shows that the Lagrangian cannot
be generic.

From a practical point of view, it is fairly easy to write down a
version of the Pyramid Scheme, which realizes these principles. We
introduce a different singlet $S_i$ for each of the trianons. The R
preserving superpotential is

$$W = W_{std} + \sum \alpha_i S_i H_u H_d + \sum \beta_i S_i {\rm Tr\ }{\cal
T}_i \bar{\cal T}_i , $$ where $W_{std}$ contains the Yukawa
couplings of the MSSM, and some of the coefficients $\alpha_i$
and/or $\beta_i$ might vanish in the R-symmetry limit. Other
polynomials in the $S_i$ are also forbidden by R-symmetry. We then
add the non-generic R-violating superpotential,
$$\delta W = W_0 + \sum M_i^2 S_i + \sum m_i {\rm Tr\ }{\cal T}_i \bar{\cal
T}_i .$$  In general, the $S_i$ might have different R charges, and
some of the $\beta_i$ might be zero.  Indeed, it turns out that the
phenomenological requirement that the R symmetry forbid dimension
four and five operators that violate $B$ and $L$ forces some of the
$\beta_i$ to vanish.  A more fundamental reason for this to be the
case is that if the $M_i$ are nonzero, and one of the $\beta_i$
vanishes, then the model has no SUSY preserving vacuum in the $m_P
\rightarrow\infty$ limit. It can thus serve as the low energy
effective field theory of a finite quantum theory of dS space.

\subsection{Dark matter and axions in the shade of the Pyramid}

There are a variety of scenarios for Dark Matter in the Pyramid
Scheme. The dynamics of the model is controlled by the ratios of the
trianon masses $m_i$ to scales determined by the gauge coupling at
the apex of the pyramid. We consider $m_{1,3}$ to be in the 10s of
TeV range, and imagine that the gauge coupling at the unification
scale is strong enough that the confinement scale $\Lambda_3$ of the
$N_F = N_C = 3$ gauge theory below these masses is a few
TeV\footnote{This appears to lead to a Landau pole below the
unification scale, which might indicate that $SU(3)$ must be the
Higgsed remnant of a larger gauge group.}.  The $1,3$ trianon
numbers are then conserved quantum numbers in the low energy field
theory and reasonable primordial asymmetries can guarantee that the
lightest particle carrying these quantum numbers is a dark matter
candidate. If this is the origin of dark matter, then searches for dark matter annihilation signals will be fruitless. Alternatively, even without asymmetries, non-thermal production of these {\it pyrma-baryons} can produce the right dark matter density.  In the latter case, dark matter will
annihilate primarily into the pseudo-Nambu Goldstone boson associated with
spontaneous breakdown of the baryonic quantum number associated with the
second trianon, whose mass is of order the confinement scale. The
mass of this PNGB is of order an MeV, so it decays into leptons and
photons only. This might be of use in explaining the results of the PAMELA and Fermi observations of excesses in the cosmic abundance of leptons.
However, it seems likely that there are also astro-physical explanations for these excesses.

Briefly then, the hidden sector of the Pyramid scheme seems rich enough
to produce a dark matter candidate in various ways. Alternatively one might imagine that the QCD axion is the dark matter. Here, an interesting issue arises. The terms in the low energy Lagrangian of the Pyramid Scheme, which preserve the discrete R symmetry, are such that one can simultaneously eliminate all phases in Yukawa couplings except that of Cabibbo-Kobayashi and Maskawa, as well as the $\theta$ parameters of both QCD and the strong group at the apex of the Pyramid. The Lagrangian is CP invariant, up to the CKM phase.  In the field basis where all other couplings are real, the terms that violate discrete R symmetry are generically complex. Recall however that these terms are non-generic.  One might try to argue, that the combination of IR CP conservation, and the thermal nature of the states on the horizon, which are putatively responsible for generating the R violating terms, implies that all CP violation in these terms should be proportional to the Jarlskog invariant.  This could solve the strong CP problem in a novel way, without the need for a QCD axion, vanishing up quark mass, or Nelson-Barr
mass matrix.

Carpenter {\it et. al.}\cite{cdf} have recently studied the QCD axion solution of the strong CP problem in the context of gauge mediated SUSY breaking. They found that the idea was strongly constrained and the constraints are not satisfied in the Pyramid Scheme. Furthermore, in order to make their models ``natural", these authors had to invoke a landscape of states with a variety of values for low energy parameters, with axion dark matter an {\it a priori} selection parameter. There are many ways in which these considerations conflict with the underlying idea of CSB. Thus, axion dark matter does not appear to be a likely prediction of the Pyramid Scheme.

We will not delve further into the properties of the Pyramid
Scheme. Its phenomenology is quite interesting, and has been
outlined in \cite{pyrma12} but a number of questions require more of
a solution of the strongly coupled gauge theory than is currently
available.

\section{Executive summary}

The holographic approach to space-time geometry views geometry as a
collective description of a network of interlocking quantum systems.
At the end of the day, processes accessible to a single observer can
all be described by an ordinary quantum Hamiltonian in a single
Hilbert space, but the form of the Hamiltonian is constrained by the
picture of overlapping systems with partial information. The only
full mathematical solution of these constraints is the dense black
hole fluid (DBHF) cosmology of \cite{holocosm}.

The quantum variables of holographic space-time are the orientations
of pixels on the holographic screen of a causal diamond. In the
limit of large diamonds they are best thought of in terms of the
degrees of freedom of relativistic super-particles, penetrating the
screen. This is the way that particle physics emerges from the
formalism, and it always emerges together with (approximate)
supersymmetry. The commutation relations of these variables define
the non-commutative geometry of compact dimensions.  The details
have only been worked out for maximally supersymmetric
compactifications. The automatic emergence of supersymmetry, and the
unification of particle statistics with the non-commutative geometry
of finite area holographic screens, are two of the most interesting
aspects of the formalism. The description of particles in terms of
block diagonal matrix algebras, raises the question of what the
off-block diagonal matrix elements mean. In our quantum theory of de
Sitter space, they represent degrees of freedom outside the horizon
of any given observer.

The DBHF cosmology does not actually contain any
observers\footnote{Observer $\equiv$ isolated quantum system with
many quasi-classical operators.}. All of its degrees of freedom are
in constant equilibrium with each other.  An heuristic description
of a more realistic cosmology can be based on the idea of low
entropy defects in the DBHF. This is a partial explanation of why
our universe began in a low entropy state, so that it could exhibit
a second law of thermodynamics.  The most probable initial
conditions for such a cosmology is one in which the defect involves
the smallest number of degrees of freedom that is compatible with
whatever {\it a priori} environmental constraints one wishes to
impose.  If we apply the Israel junction condition to the asymptotic
future state of the defect, embedded in the DBHF, we conclude that
it must approach a dS space, with the largest c.c. compatible with
the {\it a priori} constraints.

The theory of particle physics is then equivalent to the quantum
theory of dS space, with a fixed, small, value of the c.c.   That
theory must approach an S-matrix theory for a super-Poincare
invariant, discrete R-symmetric, model. The SUSY breaking scale is
related to the c.c. by the formula $m_{3/2} = K \Lambda^{1/4}$,
where the current best estimate of $K$ is $o(10)$.  For very small
$\Lambda$ this situation must be described approximately by
effective field theory, and this implies that the dS space is four
dimensional. This is the only dimension in which SUGRA has dS
solutions.  A further constraint is that the CDL tunneling
probability from the dS minimum to any negative energy density
region of the effective potential, must behave as $e^{ - \pi
(RM_P)^2 }$, in the limit as $RM_P \rightarrow\infty$. This rules
out models of SUSY breaking based on flat space field theory at
meta-stable states.  When we combine old results of Nelson and
Seiberg with the R symmetry properties of CSB, we come to the
conclusion that the low energy Lagrangian cannot be the most generic
form consistent with some set of symmetries.

In the framework of CSB, the R invariant part of the Lagrangian is
that of a super-Poincare invariant S-matrix theory, much like
ordinary string theory. Experience shows that, in both asymptotically flat  and AdS string theory, such Lagrangians
satisfy the usual rules of genericity, familiar from the
renormalization group in quantum field theory. In dS space on the other hand, the R violating parts of the Lagrangian, come from interactions between particle-like degrees of freedom in a given horizon volume, and the degrees of freedom on the horizon.  They all come from diagrams where a single
gravitino line goes out to the horizon and returns.  Higher order
diagrams are exponentially suppressed as the c.c. goes to zero.
There is thus no reason to assume that the terms in the R breaking
Lagrangian are generic.

It has turned out to be extraordinarily difficult to construct a low
energy Lagrangian consistent with all of these constraints, with the
rudiments of low energy phenomenology, and with gauge coupling
unification. This has led to a more or less unique model, dubbed The
Pyramid Scheme, since it has a pyramidal/tetrahedral quiver diagram.
Aspects of this model should be testable at the LHC.

There are two big lacunae in the framework for particle physics
outlined here. The first is the classification of compactifications
of the holographic formalism, which lead to minimal SUSY in four
dimensions. One would like to know what constraints this puts on the
four dimensional gauge group.  Recall that conventional string
theory has produced no examples of exactly $N=1$ super-Poincare
invariant models in four dimensions, with a compact moduli space. In
the holographic formalism it is clear that for finite c.c. there are
no continuous moduli (everything is determined by the representation
theory of compact finite dimensional super-algebras), so things
should be even more constrained.

The other unsolved problem is to write down dynamical equations for
the scattering matrix.  This is of course not a new problem.
Perturbative string theory gives us an asymptotic series for the
scattering matrix, but, except for those special cases where Matrix
Theory is applicable ($\leq 4$ out of $11$ compactified dimensions)
we do not have a non-perturbative formulation of the theory. In no
case do we have a non-perturbative formulation of models of quantum
gravity in asymptotically flat space, which manifests all of the
symmetries of the system.

In ancient times, the analytic S-matrix program attempted to find a
set of equations that completely determined the scattering matrix
without referring to the bulk of space-time. This program failed
because the high energy behavior of amplitudes was not under
control. In a perturbative context, with a finite number of stable
particles, the analytic S-matrix rules seem to lead back to quantum
field theory, with the usual issues of renormalization replaced by
the question of subtractions in dispersion relations. Recently, we
have seen the development of new rules for direct computation of the
scattering matrix, particularly in maximally supersymmetric
models\cite{newrules}. In my opinion, the best way forward is to try
to work out the analogous formalism for an S-matrix for eleven
dimensional SUGRA.  This model has the maximal space-time symmetry
and simplest particle content, among all possible super-Poincare
invariant S matrices including gravitons. I would guess that once we
have a completely non-perturbative formulation of this model in
hand, it will be relatively straightforward to generalize it to
compactified models with less SUSY.  The combination of these
dynamical equations and the kinematic classification of quantum
theories of dS space, which we described above, will give us the
tools to construct a real theory of the world we live in, and to
understand the extent to which ``God had no choice in its creation".

On a more pedestrian level, there are a number of issues with the low energy Pyramid Scheme, which remain to be worked out.  Primary among these is the development of tools for calculating the spectrum of superpartners, and their interactions. The model will stand or fall on its comparison with LHC data, but we need more precise predictions in order to make this comparison.
The issue of Landau poles in the Pyramid gauge coupling also needs work. In \cite{pyrma12} we suggested embedding  $SU_P(3)$ in $SU_P(4)$, but did not provide a dynamical implementation of the Higgs mechanism. Another avenue for resolving this problem is suggested by the vanishing one loop beta function of SUSY QCD with $9$ flavors and $3$ colors. Perhaps the multidimensional RG flows in the space of this gauge coupling and the Yukawa couplings, could lead to some sort of quasi fixed point, which eliminates the Landau pole below the unification scale. Preliminary investigation of this scenario has not had promising results, but more work remains.

Finally, it would be interesting to find a connection between the supersymmetric limit of the Pyramid Scheme, and D-brane constructions in string theory. In this connection, M. Cvetic has pointed out to me that $3$ $D3$ branes at the $Z_3$ orbifold, provide a local model of the base of the Pyramid, including the $3$ generations of standard model matter. It is of interest to see if one can obtain a local model of the full Pyramid by adding D7 branes to the mix.

\vskip.4in\section{\bf Acknowledgments}

I would like to thank Willy Fischler and Tomeu Fiol for important
input into the program described in this note.  This research was
supported in part by DOE grant number DE-FG03-92ER40689.
\vfill\eject

\end{document}